\documentclass[]{spie}  

\usepackage{amsmath,amsfonts,amssymb}
\usepackage{graphicx}
\usepackage[colorlinks=true, allcolors=blue]{hyperref}
\usepackage{float}
\usepackage{pifont}

\title{Characterisation and assessment of the SOXS Spectrograph UV-VIS Detector System}

\author[a,b  ]{   Rosario Cosentino             }
\author[a    ]{   Marcos Hernandez               }
\author[a    ]{   Hector Ventura                }
\author[c    ]{   Sergio Campana                }
\author[d    ]{   Riccardo Claudi               }
\author[e    ]{   Pietro Schipani               }
\author[c    ]{   Matteo Aliverti               }
\author[c    ]{   Laura Asquini                 }
\author[d    ]{   Andrea Baruffolo              }
\author[d    ]{   Federico Battaini             }
\author[f,g  ]{   Sagi Ben-Ami                  }
\author[f    ]{   Alex Bichkovsky               }
\author[e    ]{   Giulio Capasso                }
\author[i    ]{   Francesco D'Alessio           }
\author[c    ]{   Paolo D'Avanzo                }
\author[f    ]{   Ofir Hershko                  }
\author[j,k  ]{   Hanindyo Kuncarayakti         }
\author[c    ]{   Marco Landoni                 }
\author[b    ]{   Matteo Munari                 }
\author[l    ]{   Giuliano Pignata              }
\author[n    ]{   Adam Rubin                    }
\author[b,o  ]{   Salvatore Scuderi             }
\author[i    ]{   Fabrizio Vitali               }
\author[p    ]{   David Young                   }
\author[q    ]{   Jani Achren                   }
\author[t    ]{   Jose Antonio Araiza-Duran     }
\author[s    ]{   Iair Arcavis                  }
\author[m,t  ]{   Anna Brucalassi               }
\author[f    ]{   Rachel Bruch                  }
\author[d    ]{   Enrico Cappellaro             }
\author[e    ]{   Mirko Colapietro              }
\author[e    ]{   Massimo Della Valle           }
\author[b    ]{   Rosario Di Benedetto          }
\author[d    ]{   Simone Di Filippo             }
\author[e    ]{   Sergio D'Orsi                 }
\author[f    ]{   Avishay Gal-Yam               }
\author[c    ]{   Matteo Genoni                 }
\author[j,k  ]{   Jari Kotilainen               }
\author[u    ]{   Gianluca Li Causi             }
\author[e    ]{   Laurent Marty                 }
\author[j    ]{   Seppo Mattila                 }
\author[f    ]{   Michael Rappaport             }
\author[d    ]{   Kalyan Radhakrishnan          }
\author[d    ]{   Davide Ricci                  }
\author[c    ]{   Marco Riva                    }
\author[d    ]{   Bernardo Salasnich            }
\author[e    ]{   Salvatore Savarese            }
\author[p    ]{   Stephen Smartt                }
\author[b    ]{   Ricardo Zanmar Sanchez        }
\author[v    ]{   Maximilian Stritzinger        }
\author[n    ]{   Matteo Accardo                }
\author[n    ]{   Leander H. Mehrgan            }
\author[n    ]{   Derek Ives                    }
\affil[a]{INAF - Fundaci\'{o}n Galileo Galilei, Bre\~{n}a Baja, Spain}
\affil[b]{INAF - Osservatorio Astrofisico di Catania, Catania, Italy}
\affil[c]{INAF - Osservatorio Astronomico di Brera, Merate, Italy}
\affil[d]{INAF - Osservatorio Astronomico di Padova, Padua, Italy}
\affil[e]{INAF - Osservatorio Astronomico di Capodimonte, Naples, Italy}
\affil[f]{Weizmann Institute of Science, Rehovot, Israel} 
\affil[g]{Harvard-Smithsonian Center for Astrophysics, Cambridge, USA}
\affil[h]{Max-Planck-Institut für Extraterrestrische Physik, Garching, Germany}
\affil[i]{INAF - Osservatorio Astronomico di Roma, Rome, Italy}
\affil[j]{Tuorla Observatory, Department of Physics and Astronomy, University of Turku, Turku, Finland}
\affil[k]{FINCA - Finnish Centre for Astronomy with ESO, Turku, Finland}
\affil[l]{Instituto de Alta Investigaci\'{o}n, Universidad de Tarapac\'{a}, Arica, Chile}
\affil[m]{Universidad Andres Bello, Santiago, Chile}
\affil[n]{European Southern Observatory, Garching, Germany}
\affil[o]{INAF - Istituto di Astrofisica Spaziale e Fisica Cosmica, Milano, Italy}
\affil[p]{Queen's University Belfast, Belfast, UK}
\affil[q]{Incident Angle Oy, Turku, Finland}
\affil[r]{Centro de Investigaciones en Optica A. C., León, Mexico}
\affil[s]{Tel Aviv University, Tel Aviv, Israel}
\affil[t]{INAF-Osservatorio Astrofisico Arcetri, Firenze, Italy}
\affil[u]{INAF - Istituto di Astrofisica e Planetologia Spaziali, Rome , Italy}
\affil[v]{Aarhus University, Aarhus, Denmark}
\authorinfo{Further author information: (E-mail: cosentino@tng.iac.es, Telephone: +34 922433642\\ }
 
\begin{document} 
\maketitle

\clearpage
\begin{abstract}
The SOXS spectrograph, designed for the ESO NTT telescope, operates in both the optical (UV-VIS: 350-850 nm) and NIR (800-2000 nm) bands. This article provides an overview of the final tests conducted on the UV-VIS camera system using a telescope simulator. It details the system's performance evaluation, including key metrics such as gain, readout noise, and linearity, and highlights the advancements made in the upgraded acquisition system. The testing process, conducted in the Padua laboratory, involved comprehensive simulations of the telescope environment to ensure the results closely resemble those expected at the ESO-NTT telescope. The successful completion of these tests confirms the system's readiness for deployment to Chile, where it will be installed on the NTT telescope, marking a significant milestone in the SOXS project.
\end{abstract}

(Ref. ~\citenum{soxscosentino,cosentino2020,cosentino2022,colapietro2024,kalyan2024,claudi2024,ricci2024}).




\section{The SOXS UV-VIS arm}
The characterization and assessment of the UV-VIS detector system were conducted utilizing SOXS UV-VIS spectrograph, installed in the telescope simulator. The telescope simulator, located in the Padua laboratory, incorporates a mechanical derotator and emulates the electronic environment, including racks, harnessing, and power supply, mirroring the final configuration anticipated for installation on the ESO-NTT telescope. The support electronics and software environment mirror the final version of the instrument to ensure test results be comparable with the expected outcomes at the telescope. Various readout modes were tested, and key characteristics including gain, readout noise, and linearity of the detector were measured.

 \begin{figure} [H]
   \begin{center}
   \begin{tabular}{c} 
   \includegraphics[height=11 cm]{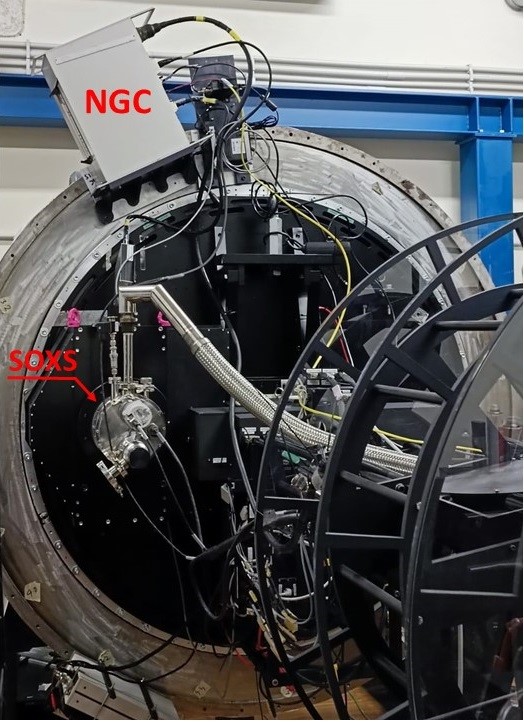}
   \end{tabular}
   \end{center}
   \caption[VIS Camera] 
   { \label{fig1} SOXS UV-VIS mounted in the telescope simulator.}
   \end{figure}

\section{Overview of the UV-VIS Detector System}
The UV-VIS detector system comprises an e2v CCD 44-82 sensor, a custom detector head integrated with the ESO CFC cooling system, the NGC CCD controller engineered by ESO, and a software interface for managing data acquisition. (Figure ~\ref{fig3}).

   \begin{figure} [H]
   \begin{center}
   \begin{tabular}{lcc} 
   \includegraphics[height=7 cm]{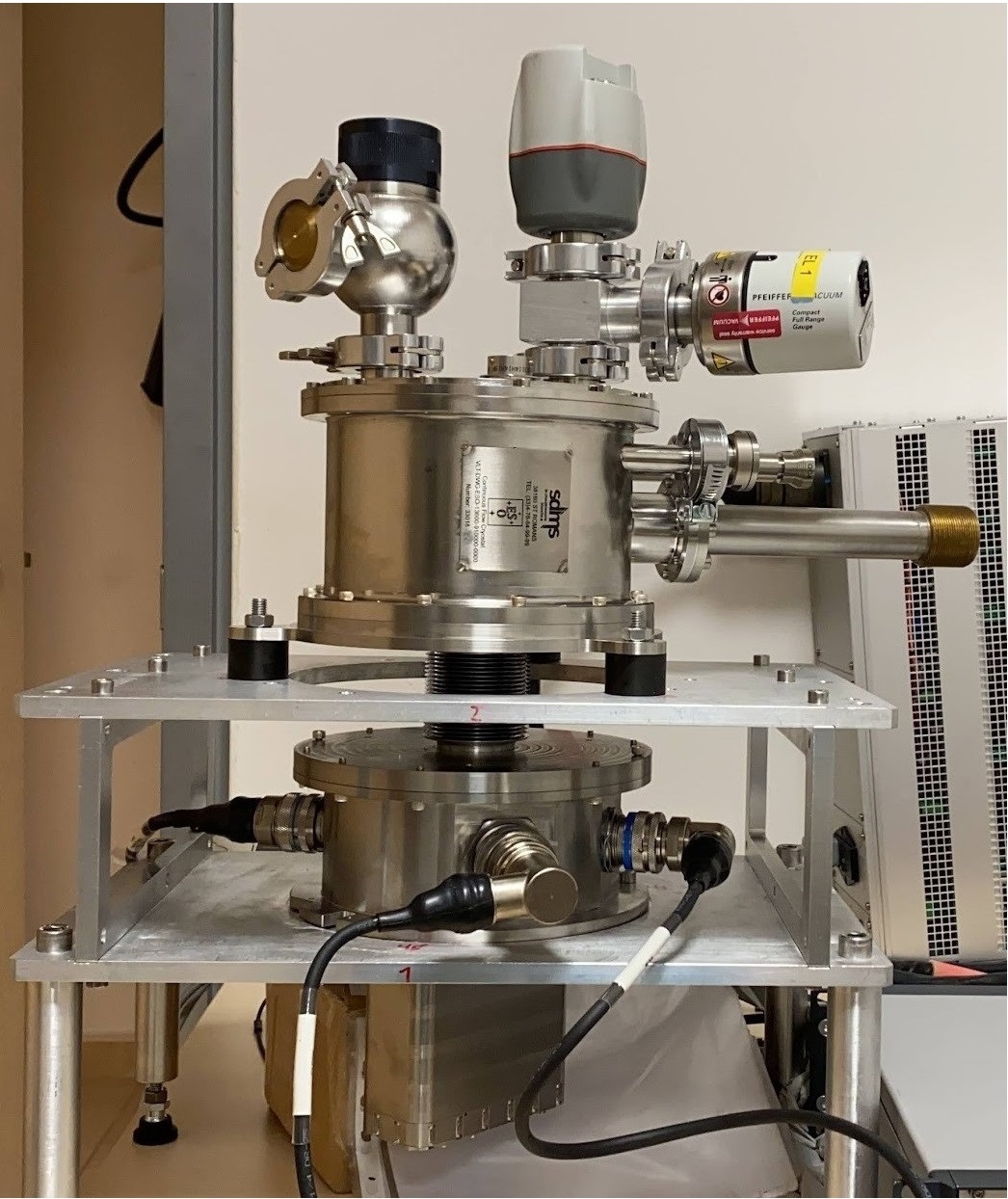}
   \includegraphics[height=7 cm]{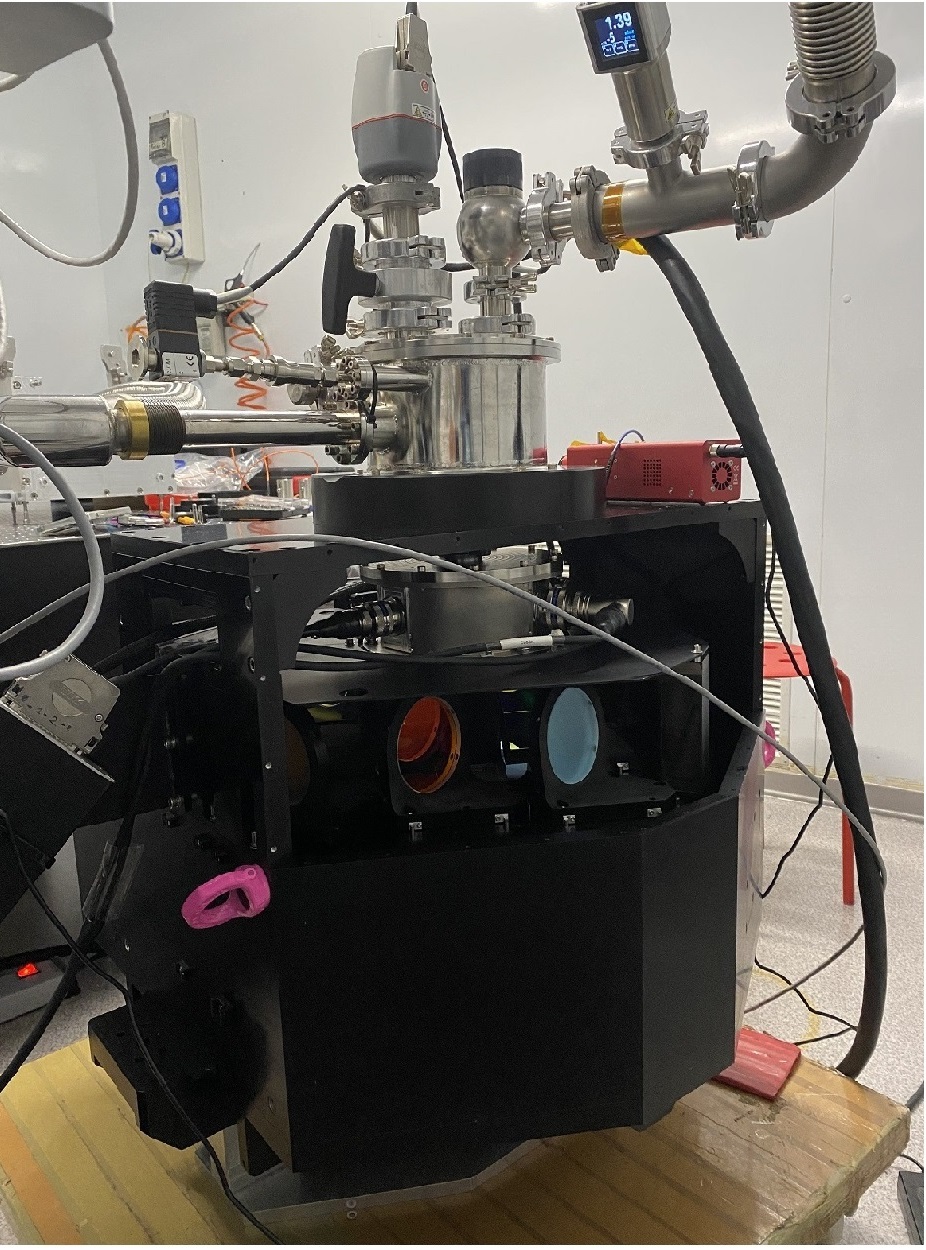}
   \end{tabular}
   \end{center}
   \caption[VIS Camera] 
   { \label{fig2} The UV-VIS Detector housing (left) and The UV-VIS Spectrograph with the Detector housing (right). }
   \end{figure}
   
The CCD control system is critical for maintaining the operational stability and performance of the UV-VIS detector. The e2v CCD 44-82 sensor is known for its high quantum efficiency and low noise, which are essential for capturing faint astronomical signals. The detector head ensures optimal thermal management by using the ESO CFC cooling system, which minimizes dark current and maintains the CCD at a stable, low temperature.
The NGC CCD controller provides advanced readout capabilities with multiple readout modes, allowing for flexibility in observational strategies. It supports various gains and readout speeds, enabling the optimization of signal-to-noise ratio (SNR) for different observational conditions. The software interface, designed for ease of use, allows astronomers to select image types (e.g., bias, dark, normal), configure acquisition parameters, and monitor system performance in real-time.

Figure ~\ref{fig3} illustrates the Detector user interface along with the results of an acquisition displayed in a graphical user interface (GUI). Through the user interface, it is possible to select the image type (bias, dark, normal, etc.), the readout mode with various gains and readout speeds, and other acquisition parameters.

   \begin{figure} [H]
   \begin{center}
   \begin{tabular}{c} 
   \includegraphics[height=7 cm]{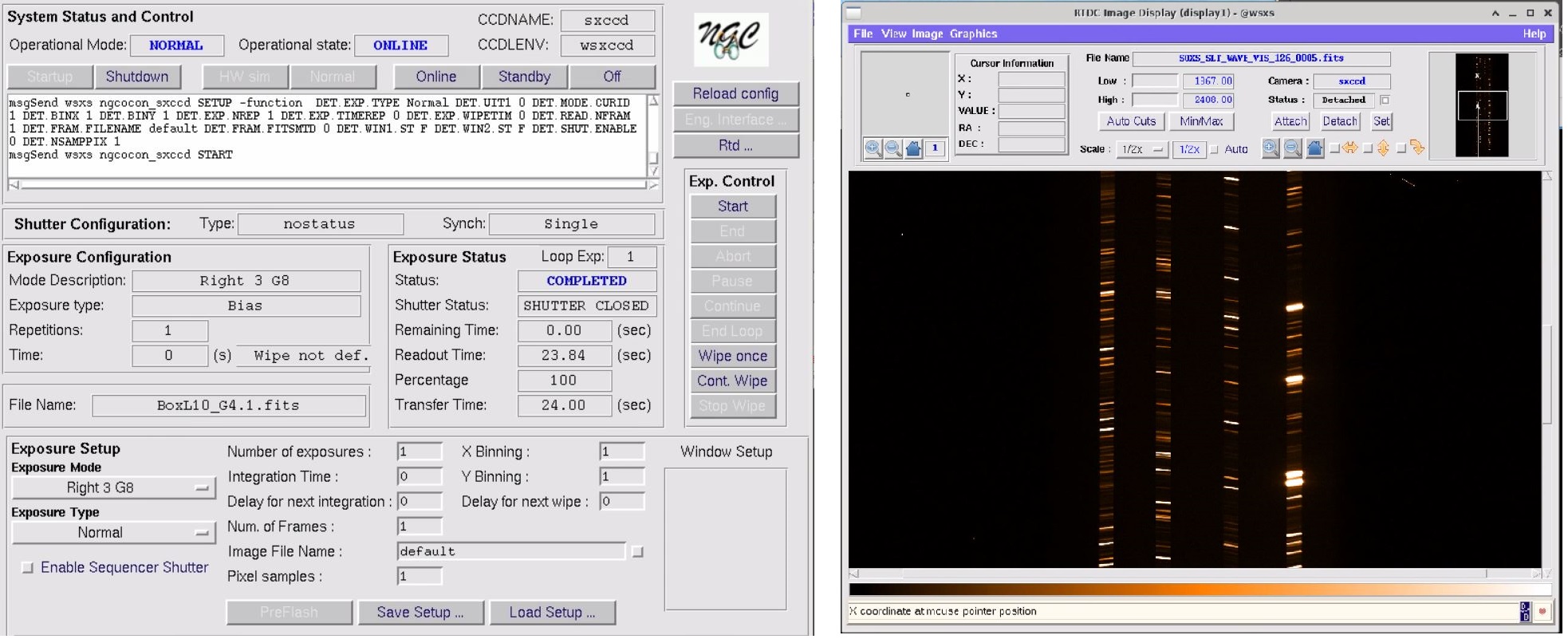}
   \end{tabular}
   \end{center}
   \caption[VIS Camera] 
   { \label{fig3} Detector user interface along with the results of an acquisition displayed in a graphical user interface (GUI).}
   \end{figure}

\section{Testing and Performance Evaluation of the SOXS UV-VIS Spectrograph}

These tests were conducted in Padua, leveraging the advanced infrastructure of the SOXS spectrograph and NGC (Next Generation Controller) integrated into the telescope interface (refer to Figure ~\ref{fig2}). The interaction between these components was facilitated by the SOXS Instrument Control System (ICS), ensuring precise and coordinated functionality throughout the testing process.
The utilization of the SOXS spectrograph and NGC within the telescope interface provides a comprehensive simulation environment, replicating the the interaction between the detector system and the observational setup. This simulation extends beyond mere hardware integration, and includes  the complexity of data acquisition, processing, and control, mirroring the operational conditions expected during actual astronomical observations.
The results obtained from these tests were compared with previous assessments conducted at the FGG (Fondazione Galileo Galilei) laboratory, providing a robust benchmark for performance evaluation. This comparative analysis ensures continuity and consistency in assessing the system's functionality and performance across different testing environments.
Furthermore, the readout values obtained across various operational modes were documented, providing valuable insights into the system's behavior under different configurations. These values serve as crucial reference points for calibrating the system and optimizing its performance for specific observational objectives.
In summary, the comprehensive testing regimen conducted in Padua, utilizing the SOXS spectrograph and NGC within the telescope interface and operated via the SOXS ICS, provided valuable insights into the system's performance characteristics. The comparative analysis with previous assessments and detailed documentation of readout values further enhance our understanding of the system's capabilities and readiness for astronomical observations.

The following sub paragraphs provide detailed information on the main characteristics of the readout modes implemented in the system, which can be used to choose the best configuration to meet astronomical needs under different observing conditions.

\subsection{Detailed Characteristics of Readout Modes for Optimal Astronomical Configurations}
This section provide detailed information on the main characteristics of the readout modes implemented in the system, which can be used to choose the best configuration to meet astronomical needs under different observing conditions.
Table 1 outlines the gain, readout noise, and readout time for each readout mode, helping in the selection of the optimal setup for various astronomical applications.

\begin{table}
 \begin{center}
  \caption[rdnoise]
  {\label{tab1}gain, readout noise, and readout time for each readout mode}
  \includegraphics[height=8 cm]{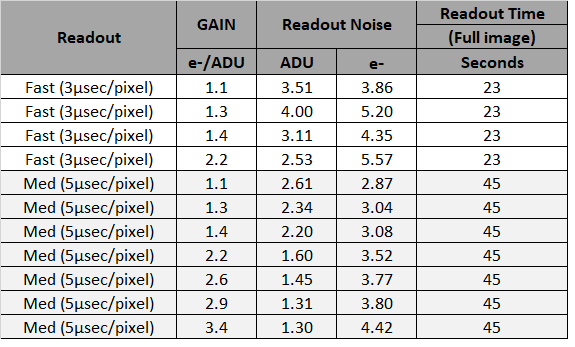}
  \end{center}
\end{table}

Table ~\ref{tab2} displays the dynamic range of the CCD (Charge-Coupled Device) along with the various gain settings that can be used. To optimize the observation of an object, the appropriate configuration will be selected based on the object's brightness. This selection process ensures that the CCD captures the best possible image by adjusting its sensitivity according to the light intensity of the object being observed.

\begin{table} [hbt!]
  \begin{center}
  \caption[dinrange]
  {\label{tab2}CCD dynamic range covered depending on gain}
  \includegraphics[height=8 cm]{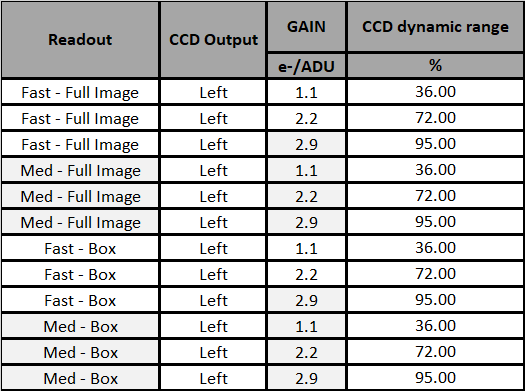}
  \end{center}
\end{table}

\subsection{CCD Image Readout Methods and Their Impact on Readout Speeds}

The acquisition system implements various methods for reading CCD images, including full image readout, selected box readout, and different binning configurations of the full image. The readout speeds vary depending on the CCD area selected for reading. Table ~\ref{tab3} and Table ~\ref{tab4} lists the readout speeds depending on the CCD area selected for reading, offering insights into how different configurations affect readout time.

\begin{table} [hbt!]
\centering
  \caption[speed3]
  {\label{tab3} readout time for Fast speed mode}
 \includegraphics[height=3 cm]{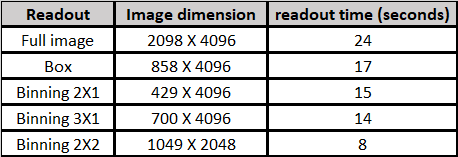}
  \end{table}

\begin{table} [hbt!]
\centering
  \caption[speed5]
  {\label{tab4} readout time for Med speed mode }
 \includegraphics[height=3 cm]{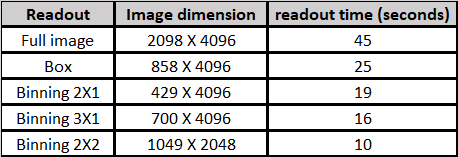}
  \end{table}

 \begin{figure} [H]
   \begin{center}
   \begin{tabular}{c} 
   \includegraphics[height=9 cm]{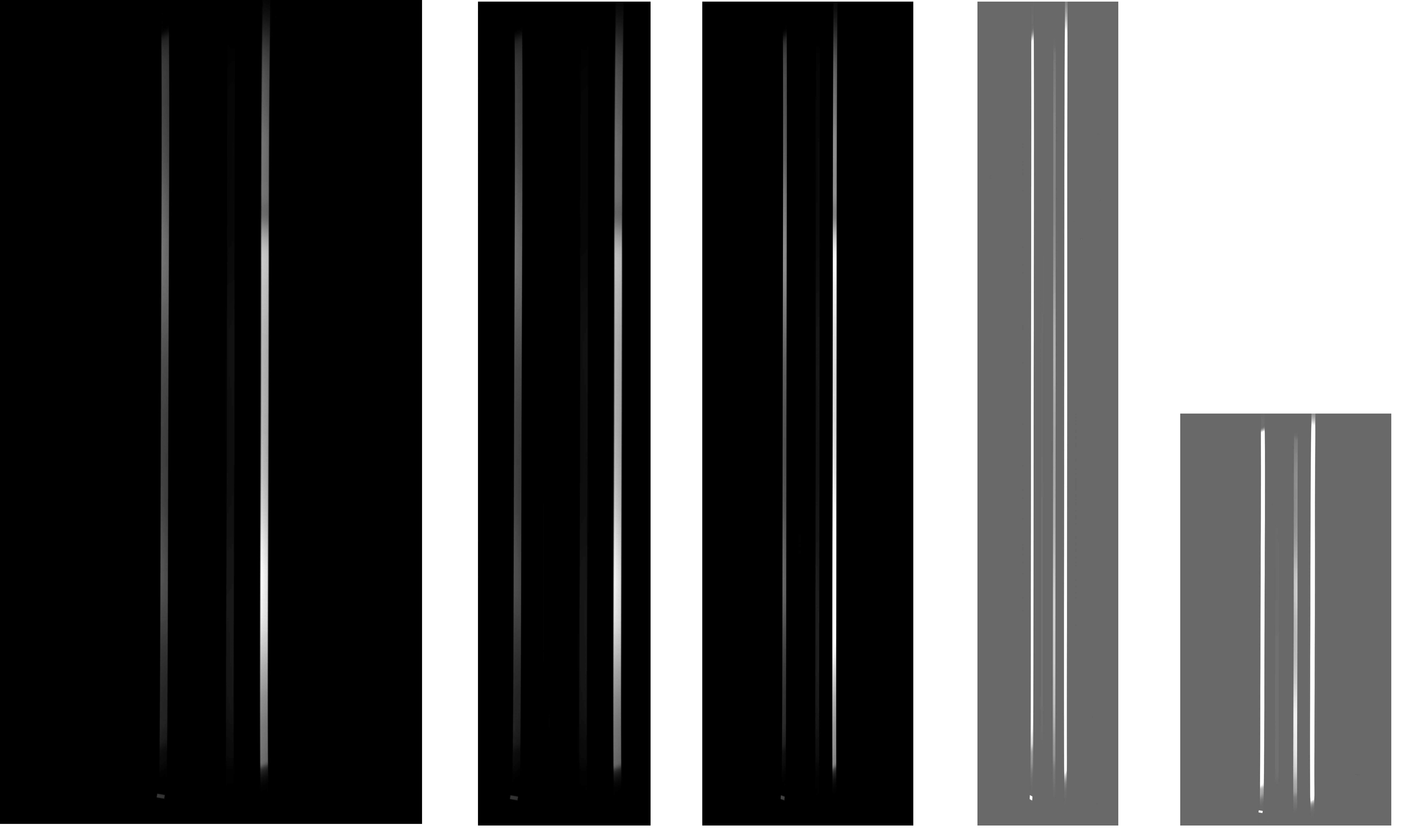}
   \end{tabular}
   \end{center}
   \caption[binning] 
   { \label{fig4} Full image, Box and binnings of the CCD.}
   \end{figure}

\subsection{Linearity}

\begin{figure} [H]
   \begin{center}
   \begin{tabular}{c} 
   \includegraphics[height=9 cm]{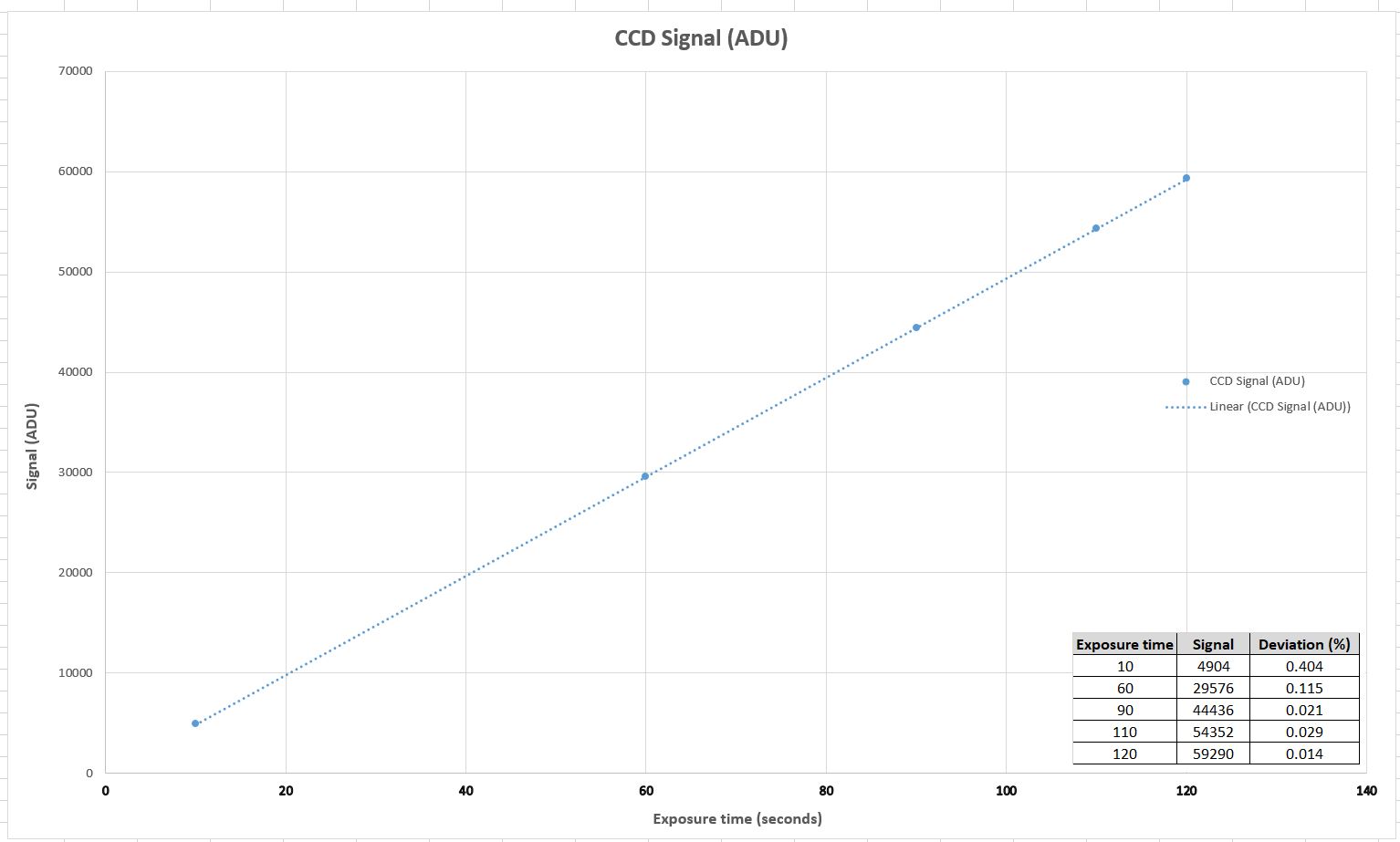}
   \end{tabular}
   \end{center}
   \caption[linearity] 
   { \label{fig5} "Linearity and deviation of the CCD across the entire dynamic range.".}
   \end{figure}

Figure ~\ref{fig5} displays the linearity and deviation from linearity of one of the readout modes implemented in the system. The measurement of CCD linearity was conducted by capturing images at different exposure times and plotting the average pixel value against the exposure time. A linear regression was then performed to determine the linearity and identify any deviations from the expected linear response. The results show that the linearity meets the requirements across the dynamic range of the CCD sensor, ensuring accurate and reliable performance in various operating conditions.

\subsection{Final Testing and Validation Using Real Spectra}

\begin{figure} [H]
   \begin{center}
   \begin{tabular}{c} 
   \includegraphics[height=9 cm]{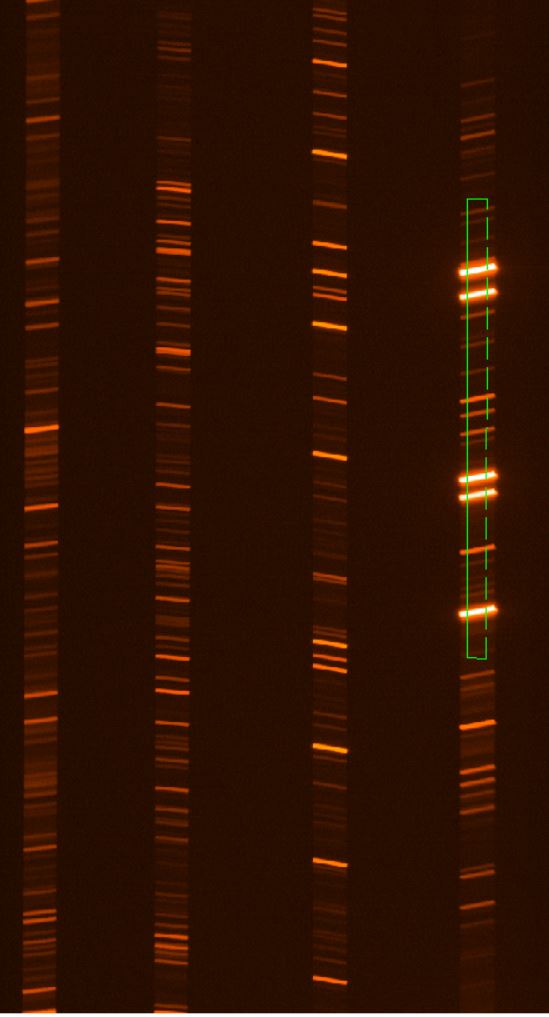}
   \end{tabular}
   \end{center}
   \caption[calibration] 
   { \label{fig6} "Real spectra of UV-VIS Spectrograph on CCD (RAW image).".}
   \end{figure}

\begin{figure} [H]
   \begin{center}
   \begin{tabular}{c} 
   \includegraphics[height=9 cm]{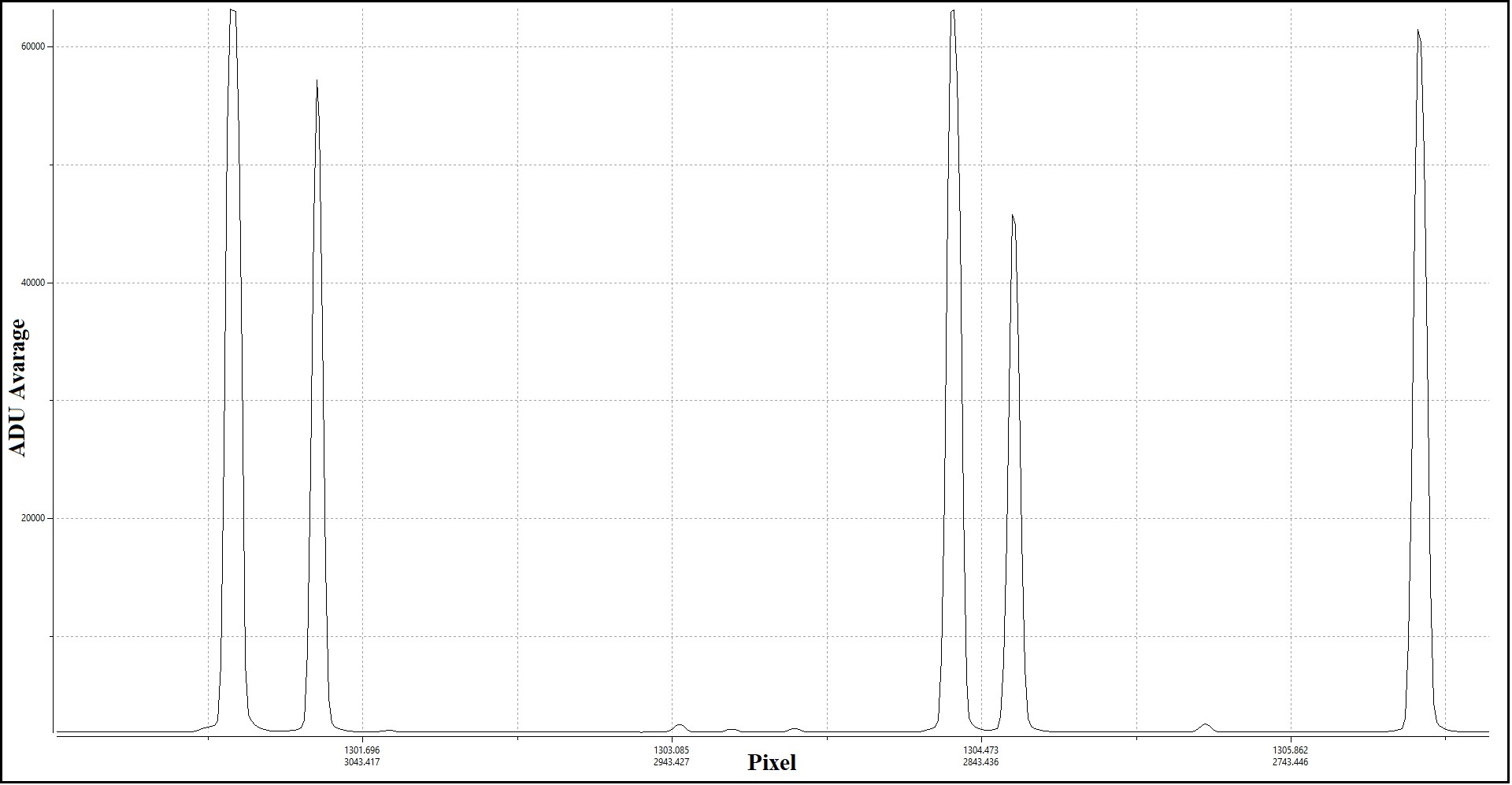}
   \end{tabular}
   \end{center}
   \caption[spectra] 
   { \label{fig7} "Real spectra of UV-VIS Spectrograph  (extracted in the green box of Figure ~\ref{fig6}.".}
   \end{figure}

The testing process consist of series of steps aimed at assessing the quality of data produced by the spectrograph and CCD system. Central to this process was the capture of calibration images, crafted to illuminate specific facets of system performance. These calibration images serve as tools for identifying and mitigating systematic errors or artefacts inherent to the detector system. By analysing the characteristics of these images, astronomers can discern subtle irregularities in the data, enabling targeted adjustments to enhance overall data quality.
Moreover, the testing process involved the extraction and analysis of spectra, providing a direct measure of the system's spectral resolution and sensitivity.  By extracting and scrutinizing these spectra, astronomers can obtain valuable information about the target object's physical properties and chemical composition. This comprehensive analysis not only validates the accuracy and fidelity of the acquired data but also provides valuable insights into the system's performance under real-world observational conditions.
The figure presented alongside showcases an illustrative example of this testing process, juxtaposing a calibration image (Figure ~\ref{fig6}) with the corresponding extracted spectrum (Figure ~\ref{fig7}). This visual representation offers a compelling demonstration of the system's capability to faithfully capture and analyse spectral data. 
In summary, the testing process employed a multifaceted approach to assess the quality of data acquired by the spectrograph and CCD system. By capturing calibration images and extracting spectra, we evaluated the system's performance and ensure the accuracy and fidelity of the acquired data. This testing regimen serves as a crucial validation step, confirming the system's readiness for astronomical observations.

\subsection{Conclusions}
The tests conducted with the UV-VIS arm of the SOXS spectrograph mounted on the telescope simulator and with the support electronics that will be installed at the telescope in Chile have functioned correctly and as expected. We measured the parameters characterizing the acquisition system, such as gain, readout noise, and linearity, and obtained the initial images and extracted the first spectra. The system is ready to be shipped to Chile and installed on the NTT telescope.
The successful characterization and testing of the UV-VIS detector system confirm that it meets the stringent requirements for astronomical observations. The measured performance metrics, including gain, readout noise, and linearity, align with the design specifications, ensuring high-quality data acquisition. The readiness of the system for deployment marks a significant milestone in the SOXS project, paving the way for groundbreaking observations at the ESO NTT telescope.
In addition to validating the core performance parameters, the final tests demonstrated the robustness of the upgraded acquisition system. The enhanced software interface and flexible readout options provide astronomers with versatile tools for optimizing observations under varying conditions. The ability to emulate the electronic and mechanical environment of the ESO-NTT telescope within the Padua laboratory was crucial in achieving these results, ensuring that the system's behavior in the field will closely match the simulated conditions.
Overall, the comprehensive testing regimen, from initial calibration to spectral extraction, has thoroughly vetted the UV-VIS arm's capabilities. The system's compliance with the design requirements and its successful performance in a controlled, simulated environment underscore its readiness for the challenging observational tasks ahead. With the deployment to Chile imminent, the SOXS spectrograph is poised to contribute significantly to our understanding of the universe through precise and reliable data collection.

\acknowledgments 
 
A special acknowledgement to the European Southern Observatory for the support provided and for the availability to share its knowledge and to allow for the use of the ESO laboratories in Garching. 


 


\bibliographystyle{spiebib} 
\bibliography{VIS} 
\end{document}